\documentclass[twocolumn,aps,prb,showpacs,amsmath,superscriptaddress]{revtex4-1}
\usepackage{graphicx}
\usepackage{epstopdf}
\usepackage{color}
\begin{document}

\title{Numerical Approach for the Evolution of Spin-Boson Systems and its Application to the Buck--Sukumar Model}
\author{Xueying Liu}
\affiliation{School of Science, Southwest University of Science and Technology, Mianyang 621010, China}
\author{Xuezao Ren}
\affiliation{School of Science, Southwest University of Science and Technology, Mianyang 621010, China}
\author{Chen Wang}
\affiliation{Department of Physics, Zhejiang Normal University, Jinhua 321004, China}
\author{Gao Xianlong}
\affiliation{Department of Physics, Zhejiang Normal University, Jinhua 321004, China}
\author{Kelin Wang}
\affiliation{Department of Modern Physics, University of Science and Technology of China, Hefei 230026, China}
\date{\today}
\begin{abstract}
    We study the evolution properties of spin-boson systems by a systematic numerical iteration approach, which performs well in the whole coupling regime. This approach evaluates a set of coefficients in the formal expansion of the time-dependent Schr\"{o}dinger equation $\vert t\rangle=e^{-i\hat{H}t}\vert t=0\rangle$ by expanding the initial state $\vert t=0\rangle$ in Fock space. This set of coefficients is unique for the studied Hamiltonian, allowing one to calculate the time evolution from different initial states. To complement our numerical calculations, we applied the method to the Buck--Sukumar model. Furthermore, when the ground-state energy of the model is unbounded and no ground state exists in a certain parameter space, the time evolution of the physical quantities is naturally unstable. The performance of the numerical method on a Hamiltonian with anti-Hermitian terms (which models open quantum systems) was also evaluated.
\end{abstract}

\maketitle

\section{Introduction}
The interaction between light and matter is a central and fundamental problem in quantum optics. The simplest light--matter interaction model is the Jaynes--Cummings (JC) model, proposed by Jaynes and Cummings within the rotation-wave approximation (RWA). The JC model describes the interaction between a two-level atom and a single-mode field~\cite{JC1962}, which is an exactly
solvable model with applications in many fields~\cite{GB1991, RE2011, HW2012, Mb2015, FS2018}. The JC model is experimentally realized
by the squeezed radiation field in a $^{85}$Rb-atom micromaser ~\cite{Rempe1987} and by the collapse and revival of atomic inversion in a $^{138}$Ba-atom microlaser ~\cite{An1994}.

The JC model has been generalized in different ways. For example, the Dicke model (also known as the Tavis--Cummings model) couples the single mode with multiple two-level atoms. The coupling of multi-modes and multi-atoms can be constructed by Yang--Baxter algebra and is integrable through the Bethe ansatz~\cite{Kundu2003}. An extra counter-rotating term in the JC model gives the famous Rabi model, which has been only analytically solved by Braak~\cite{Braak2011} till date. The JC model can also be generalized by including nonlinear couplings between a single atom and the radiation field~\cite{MH2014, FS2014, LL2018, A2012}. An example is the intensity-dependent JC
model, which exhibits significantly different dynamics in the absence and presence of RWA, owing to the dramatically enhanced field-squeezing effect in the latter case~\cite{Ng2000}.
The Buck--Sukumar (BS) model~\cite{BS1981} is a more general JC model considering the interaction of a two-level system with a field under intensity-dependent
coupling, which is still analytically solvable.
In contrast, the BS model with the counter-rotating term (sometimes called the nonlinear Rabi model) is non-analytic. Although the nonlinear term in the intensity-dependent JC or the BS model involves multiphoton interactions that are not physically realizable in current cavity- or circuit-QED setups, they are still of interest to the quantum optics and cold-atom fields. For instance, the cold atoms in optical lattices may provide a means of realizing the nonlinear light--atom coupling appearing in BS-type Hamiltonians, via the engineering of adequate optical lattices~\cite{Celi2017}.

The dynamical nature of the quantum Rabi-like model has been recently discussed. Through time-dependent correlation functions, the dynamical correlation functions of the quantum Rabi model involve the JC scheme in the ultrastrong-coupling and deep strong-coupling regimes~\cite{Wolf2013,Xie2017}. The nonlinear dynamics of trapped-ion models have been proposed for blocking the propagation of quantum information along the Hilbert space of the JC and quantum Rabi models~\cite{Cheng2017}.
Exploiting the parity symmetry of the Rabi model, Hu et al. ~\cite{Hu2017} recursively evolved the
corresponding positive and negative parities using the expansion coefficients in the dynamical equations.

In this paper, quantum spin-boson systems expressed by time-independent Hamiltonians $\hat{H}$ are evolved by a non-perturbation numerical approach. When the stationary eigenvalues of the Hamiltonians are known, the non-perturbation numerical approach recovers the existing results~\cite{HW2012, LW2009} obtained by evolving the known eigenvalues and eigenvectors of $\hat{H}$~\cite{ZX2014, Braak2011} from the initial state. However, when the eigenvalues are unknown, our method can evolve the properties from different initial states. Herein, we apply this method to the BS model and discuss the evolution of the mean photon number, the atomic inversion, and the ground-state properties in the absence and presence of RWA. Ng {\it et al.} proved that the BS Hamiltonian is non-real positive-definite when the strengths of the rotating- and counter-rotating-wave terms are equal~\cite{Ng2000}. Such a system has no physical meaning and no definite time evolution. However, Rodr\'{i}guez-Lara {\it et al.}~\cite{Rodrguez2013} temporally evolved the equivalent nonlinear JC photonic lattice using quantum optics-based methods via the analogy between the transport of single-photon states and propagation of a classical field. We resolve this obvious contradiction through numerical calculations.

The remainder of this paper is organized as follows. Section ~\ref{sec:model} introduces the BS model, and Sec.~\ref{sec:method} presents our non-perturbation numerical method. Sections~\ref{sec:dis1} and ~\ref{sec:dis2} present the BS results with and without RWA, respectively. Conclusions are presented in Sec.~\ref{sec:con}.

\section{The model}~\label{sec:model}
The BS model generalizes the JC model to allow nonlinear couplings~\cite{Ng2000}. The Hamiltonian $(\hbar=1)$ of the BS model is given by
\begin{equation}
\label{H}
\begin{split}
\hat{H} &=\omega_{f}\hat{a}^{\dagger}\hat{a}
+\frac{\omega_{0}}{2}\hat{\sigma}_{z}
+g_{-}(\hat{a}\sqrt{\hat{a}^{\dagger}\hat{a}}\hat{\sigma}_{+}
+\sqrt{\hat{a}^{\dagger}\hat{a}}\hat{a}^{\dagger}\hat{\sigma}_{-})\\
&+g_{+}(\hat{a}\sqrt{\hat{a}^{\dagger}\hat{a}}\hat{\sigma}_{-}
+\sqrt{\hat{a}^{\dagger}\hat{a}}\hat{a}^{\dagger}\hat{\sigma}_{+}),
\end{split}
\end{equation}
where $ \omega_{f}$ and $ \omega_{0}$ are the frequencies of the field and the two-level transition, respectively; $\hat{a}^\dagger (\hat{a})$ is the creation (annihilation) operator of the field; and $\hat{\sigma}_{i} $ ($ i= z,+,-$) are Pauli matrices. The nonlinear coupling is split into a rotating-wave term
described by the coupling strength $ g_{-}$ and a counter-rotating term described by the coupling strength $ g_{+}$.

When $g_+=0$, we restore the original BS model without the counter-rotating term.
In this case, the energy is lower-bound
only if $g_- \le 1/(2S)$, where $S$ denotes the eigenvalues of the constant of motion of the Casimir operator $\hat{S}^2$. In terms of the Pauli operators, $\hat{S}^2 = \hat{S}_{\mp}\hat{S}_{\pm} + \hat{S}^2_z \pm \hat{S}_z$.
When $g_- > 1/(2S)$, the eigenenergies can be arbitrarily low. As there is no stable ground state, the Hamiltonian is unbounded and no ground state exists. In this sense, the model appears to be incomplete for $g_- > 1/(2S)$. The ground-state energy decreases indefinitely when the eigenvalues of $\hat{C}=\hat{a}^\dagger\hat{a}+2\hat{S}_z$, denoted as $c$, increase above a critical value. This effect illustrates a dramatic phase transition
between the normal phase and the super-radiant phase~\cite{Cordeiro2008}.

When the counter-rotation term is included ($g_+\ne 0$), the Hamiltonian cannot be analytically diagonalized, but the eigenstates and eigenenergies can be computed by a numerical approach.
After expanding the initial state in terms of the numerically determined eigenstates and eigenenergies, one can temporally evolve the physical quantities.
The following section proposes another formal and systematic numerical method, which does not require precalculation of the stationary eigenstates and eigenenergies.

\section{Non-Perturbative Time Evolution}
\label{sec:method}

The dynamics of a Hamiltonian system are given by the time-dependent Schr\"{o}dinger equation,
\begin{equation}
i\hbar \frac{\partial}{\partial t} \vert  t\rangle=\hat{H}\vert  t\rangle,
\end{equation}
which is formally solved as
\begin{equation}\label{TaylorE}
\vert  t\rangle=e^{-i\hat{H}t}\vert 0\rangle
=\sum_{n=0}^{\infty} \frac{1}{n!}(-i\hat{H}t)^n \vert 0\rangle.
\end{equation}
Here, $\vert 0\rangle$ denotes the initial ground state $\vert t= 0\rangle$.
Defining
 \begin{equation}
 \vert B _{n}\rangle=\frac{1}{n!}(-i\hat{H}t)^n \vert 0\rangle,
 \end{equation}
we notice that $\vert B _{n}\rangle$ can be concisely iterated as
 \begin{equation}
 \label{iter-1}
 \vert B _{n+1}\rangle=-\frac{it}{n+1}\hat{H}\vert B _{n}\rangle.
 \end{equation}
$\hat{H}$ can be easily operated on the initial state $\vert 0\rangle$ when $\vert 0\rangle$ is expanded in the Fock state space $\{\vert p\rangle\}$, $p=0, 1, 2,...$ or the coherent space $\{\vert \alpha \rangle\}$.

In the following analysis, we illustrate the proposed numerical recipe on a BS model. The time-dependent wave function of the two-level system is expanded as an excited state $\vert e\rangle$ and the ground state $\vert g\rangle$, namely, as $\vert  t\rangle =\vert t\rangle_{1}\vert e\rangle+\vert t\rangle_{2}\vert g\rangle $, where $\vert t \rangle_{1}$ and $\vert t \rangle_{2}$ are the time-dependent wavefunctions in the $\vert e\rangle$ and $\vert g\rangle$ states, respectively. Furthermore, $\vert t \rangle$ is expanded as the following Taylor series:
\begin{equation}
\label{tstate}
\begin{split}
\vert t\rangle &=e^{-i\hat{H}t}(\vert t\rangle_{1}\vert e\rangle+\vert t\rangle_{2}\vert g\rangle)\\
&=\sum_{n=0}^\infty
\frac{1}{n!}(-i\hat{H}t)^n(\vert 0\rangle_{1}\vert e\rangle+\vert 0\rangle_{2}\vert g\rangle).
\end{split}
\end{equation}
Defining
 \begin{equation}
 \label{bn1}
 \vert B _{n}\rangle=\frac{1}{n!}(-i\hat{H}t)^n(\vert 0\rangle_{1}\vert e\rangle+\vert 0\rangle_{2}\vert g\rangle),
 \end{equation}
we recover the iteration relation Eq.~(\ref{iter-1}).
Expanding $ \vert B _{0}\rangle$ as $\vert 0\rangle_{1}\vert e\rangle+\vert 0\rangle_{2}\vert g\rangle $, Eq.~(\ref{bn1}) can be rewritten as the following operator iteration relation:
\begin{equation}
\label{bn3}
\vert B_{n}\rangle= \frac{1}{n!}(-i\hat{H}t)^n\vert B _{0}\rangle.
\end{equation}

Once $\vert B_{n}\rangle$ is known, the time evolution of the state is clearly given by
\begin{equation}
\label{bn3-1}
\vert t\rangle=\sum_n \vert B_n \rangle.
\end{equation}
The initial states can then be expanded in the orthogonal complete Fock basis $\{ \vert p\rangle\}$ as
\begin{equation}
\label{bn2}
\vert B _{n}\rangle= \sum_{p=0}^\infty (f_{p}^{1,n}(t)\vert e\rangle+f_{p}^{2,n}(t)\vert g\rangle)\otimes \vert p\rangle.
\end{equation}
Substituting $\vert B _{0}\rangle= \sum_{p=0}^\infty (f_{p}^{1,0}\vert e\rangle+f_{p}^{2,0}\vert g\rangle)\otimes \vert p\rangle,$ into Eq.~(\ref{bn3}),
applying $\hat{a}\vert p\rangle=\sqrt{p-1}\vert p\rangle$ and $\hat{a}^\dagger \vert p\rangle=\sqrt{p}\vert p\rangle$ (which appear in $\hat{H}^n$ over the Fock basis $\vert p\rangle$), and comparing the result with Eq.~(\ref{bn2}), we obtain the following iterative relationship between the corresponding expansion coefficients $ f_{p}^{i,n} (i=1,2; p=0,1,2,...) $,
\begin{equation}
\label{ite-10}
\begin{bmatrix} f_{0}^{1,n}(t) \\ f_{1}^{1,n}(t)\\.\\.\\.\\ f_{0}^{2,n}(t) \\ f_{1}^{2,n}(t)\\.\\.\\. \end{bmatrix} = \frac{(-it)^n}{n!} Q^n \begin{bmatrix} f_{0}^{1,0}(0) \\ f_{1}^{1,0}(0)\\.\\.\\. \\f_{0}^{2,0}(0)\\f_{1}^{2,0}(0)\\.\\.\\. \end{bmatrix},
\end{equation}
where $Q$ is the transfer matrix, given by
\begin{equation}
Q=\begin{bmatrix} Q_{11} & Q_{12} \\ Q_{21} & Q_{22} \end{bmatrix}, \quad
\end{equation}
and
\begin{equation}\label{Qg}
\begin{split}
 &Q_{11}=(\omega_{0}/2+\omega_{f}p)\delta_{p,p^{'}},\\
 &Q_{22}=(-\omega_{0}/2+\omega_{f}p)\delta_{p,p^{'}},\\
 &Q_{12}=g_{+}(p+1)\delta_{p+1,p^{'}}+g_{-}p\delta_{p-1,p^{'}},\\ &Q_{21}=g_{-}(p+1)\delta_{p+1,p^{'}}+g_{+}p\delta_{p-1,p^{'}}.
\end{split}
\end{equation}
This formulation transforms the operator iteration relations between the $\vert B _{n}\rangle $ into matrix relations between the $ f_{p}^{i,n} (i=1,2)$. As the Hamiltonian multiplication $\hat{H}^n$ is absorbed in the matrix multiplication $Q$ as $Q^n$, the number of numerical calculations is greatly reduced, gaining an additional advantage. In this formulation, the matrix multiplication $Q^n$ is uniquely determined by the Hamiltonian and is independent of the $ f_{p}^{i,0} (i=1,2; p=0,1,2,...) $ determined by the initial states. The time-dependent properties of the system under different initial states depend on the same $Q^n$. Therefore, the matrix multiplication $Q^n$ can be calculated and stored prior to defining the arbitrary initial states.

To avoid error accumulation during the time evolution, we discrete the time $t$ into $ K $ steps, where each unit period $ \Delta t $ satisfies $ t=K\Delta t $. The term
\begin{equation}
\label{mt}
M(t)=\left[\frac{(-i \Delta t)^n}{n!}Q^n\right]^{K}, K=0,1,2,...
\end{equation}
is saved after each calculation. When $ M(t) $ acts on $ f_{p}^{i,0} (i=1,2; p=0,1,2,...)$ for different $ K $, $ f_{p}^{i,n}(t) (i=1,2; p=0,1,2,...) $ at each time is obtained as
\begin{equation}
\label{Mt1}
\begin{bmatrix} f_{0}^{1,n}(t) \\ f_{1}^{1,n}(t)\\.\\.\\.\\ f_{0}^{2,n}(t) \\ f_{1}^{2,n}(t)\\.\\.\\. \end{bmatrix} = M(t) \begin{bmatrix} f_{0}^{1,0}(0) \\ f_{1}^{1,0}(0)\\.\\.\\. \\f_{0}^{2,0}(0)\\f_{1}^{2,0}(0)\\.\\.\\. \end{bmatrix}.
\end{equation}

In summary, once the initial state is given, $ f_{p}^{i,0}$, $i=1,2$; $p=0,1,2,...$ can be determined, and the subsequent $f_{p}^{i,n}(t), i=1,2; p=0,1,2,...$ can be calculated by Eqs.~(\ref{ite-10})--(\ref{Mt1}). Numerically, we must truncate $n$ and $p$. Different combinations of the truncated forms, denoted as $N$ and $P$, respectively, were tested in the present study. The results with truncation errors below $10^{-9}$ are presented below.

Now, assume a coherent initial state of the system:
\begin{equation}
\vert 0\rangle=e^{-\alpha^2/2}\sum_{p} \frac{\alpha^p}{\sqrt{p!}}\vert p\rangle (\sin\theta \vert e\rangle + \cos\theta \vert g\rangle),
\end{equation}
where the amplitude $\alpha $ of the coherent state and the mean photon number of the initial state are related: $\alpha^2=\langle p \rangle$. Given $ \vert B _{0}\rangle=\vert 0\rangle_{1}\vert e\rangle+\vert 0\rangle_{2}\vert g\rangle $ and Eq.~(\ref{bn2}), $ f_{p}^{i,0}(0)$ for $i=1, 2$ are obtained as
\begin{equation}\label{initialf}
\begin{split}
&f_{p}^{1,0}(0)=e^{-\alpha^2/2}\frac{\alpha^p}{\sqrt{p!}}\sin\theta,\\
&f_{p}^{2,0}(0)=e^{-\alpha^2/2}\frac{\alpha^p}{\sqrt{p!}}\cos\theta.
\end{split}
\end{equation}

The resulting wavefunction $ \vert t\rangle $, given by
\begin{equation}
\label{wf}
\vert t\rangle=\sum_{n}\sum_{p} (f_{p}^{1,n}(t) \vert p\rangle \vert e\rangle +f_{p}^{2,n}(t) \vert p\rangle \vert g\rangle),
\end{equation}
can be numerically calculated by the above procedure.
In the following analysis, we investigate two typical measurements of interest in the quantum optics community. The first is the mean photon number $ \langle \hat{n} \rangle $, expressed as
\begin{eqnarray}
\label{photonum}
\langle \hat{n} \rangle &=& \langle t \vert \hat{a}^{\dagger}\hat{a} \vert t\rangle \nonumber\\
&=& \sum_{m}\sum_{q}\left[f_{q}^{1,m*}(t) \langle q \vert \langle e \vert +f_{q}^{2,m*}(t) \langle q \vert \langle g \vert \right]\hat{a}^{\dagger}\hat{a} \nonumber\\
&&\times\sum_{n}\sum_{p} [f_{p}^{1,n}(t) \vert p\rangle \vert e\rangle +f_{p}^{2,n}(t) \vert p\rangle \vert g\rangle]\nonumber\\ &=&\sum_{m,n}\sum_{p}p[f_{p}^{1,m*}(t)f_{p}^{1,n}(t)
+f_{p}^{2,m*}(t)f_{p}^{2,n}(t)].
\end{eqnarray}

The second is the atomic inversion $ \langle \hat{\sigma}_{z} \rangle $:
\begin{eqnarray}
\label{inversion}
\langle \hat{\sigma}_{z} \rangle &= &\langle t \vert \hat{\sigma}_{z} \vert t\rangle \nonumber\\
&=& \sum_{m}\sum_{q}[f_{q}^{1,m*}(t) \langle q \vert \langle e \vert +f_{q}^{2,m*}(t) \langle q \vert \langle g \vert ]\nonumber\\
&&\times(\vert e\rangle\langle e \vert - \vert g\rangle\langle g \vert )\nonumber\\
&&\times\sum_{n}\sum_{p} [f_{p}^{1,n}(t) \vert p\rangle \vert e\rangle +f_{p}^{2,n}(t) \vert p\rangle \vert g\rangle]\nonumber\\ &=&\sum_{m,n}\sum_{p}[f_{p}^{1,m*}(t)f_{p}^{1,n}(t)-f_{p}^{2,m*}(t)
f_{p}^{2,n}(t)].
\end{eqnarray}
Physically, the mean photon number and the atomic inversion may periodically collapse and revive, as observed in the JC model.

For comparison, we briefly describe the time evolution from the known eigenvalues and eigenvectors of $\hat{H}$ (hereafter, this method is abbreviated as the TEEE method). The stationary state of Eq.~(\ref{H}) is assumed as $ \vert \psi\rangle=\sum_{p=0}^{\infty}[a_{n}\vert p \rangle\vert e \rangle+b_{n}\vert p \rangle\vert g \rangle $, where $ \vert p \rangle$ is the Fock state in the Fock representation. Solving $ \hat{H}\vert \psi\rangle=E\vert \psi \rangle $, we get
\begin{equation}
\label{traditional}
\begin{split}
&(\omega_{f}p+\frac{\omega_{0}}{2})a_{p}+g_{-}(p+1)b_{p+1}+g_{+}pb_{p-1}=Ea_{p},\\
&(\omega_{f}p-\frac{\omega_{0}}{2})b_{p}+g_{+}(p+1)a_{p+1}+g_{-}pa_{p-1}=Eb_{p}.
\end{split}
\end{equation}
After numerically solving the above equations, we obtain the eigenenergy spectrum $ \{E_{j}\} $ and the corresponding stationary-state set $ \{a_{p}^{j},b_{p}^{j}\} $, where $j=0,1,2,...,2P+2$ with $P$ being the truncation of the integer number $p$.
Then, expanding the given initial state $ \vert 0\rangle $ in terms of the stationary-state set $ \vert 0\rangle=\sum_{j=0}^{2P+2} F_{j}\vert  \psi_j\rangle $, where $ F_{j} $ is the expansion coefficient, the time evolution of the wavefunction is obtained as
\begin{eqnarray}\label{Eq:TEEE}
\vert t\rangle &=&e^{-i\hat{H}t}\vert  0\rangle\nonumber\\
&=&\sum_{j=0}^{\infty}F_{j}e^{-iE_{j}t}\sum_{p=0}^{\infty}\left[a_{n}\vert p \rangle\vert e \rangle+b_{n}\vert p \rangle\vert g \rangle\right]\nonumber\\
&\approx&\sum_{j=0}^{2P+2}F_{j}e^{-iE_{j}t}\sum_{p=0}^{P}\left[a_{n}\vert p \rangle\vert e \rangle+b_{n}\vert p \rangle\vert g \rangle\right].
\end{eqnarray}
As in Eqs.~(\ref{wf})--(\ref{photonum}), we can calculate the time evolutions of the mean photon number $ \langle \hat{n} \rangle=\langle t \vert \hat{a}^{\dagger}\hat{a} \vert t\rangle $ and the atomic inversion $ \langle \hat{\sigma}_{z} \rangle=\langle t \vert \hat{\sigma}_{z} \vert t\rangle $.

Here, we emphasize that both the Taylor expansion in Eq.~(\ref{TaylorE}) and the basis expansion in Eq.~(\ref{bn2}) include an infinite sum of terms. Therefore, the accuracy of the time evolution can be systematically refined by including more Taylor series and Fock states, respectively. The expansion converges quickly due to the factorial of $n$ in the denominator. This approach, which we call the non-perturbation time-evolution method, offers several advantages over expanding the eigensolutions (i.e., the TEEE method). First, this method is directly applicable when the eigenenergy and eigenstate are unknown. Second, it systematically obtains the long-time evolution. Finally, the expansion coefficients $M(t)$ in Eq.~(\ref{Mt1}) need not be recalculated in the time evolutions of different initial states.

\section{BS model with RWA $ (g_{+}=0) $}
\label{sec:dis1}
\begin{figure}[htbp]
\begin{center}
 \centering
\includegraphics[width=0.5\textwidth]{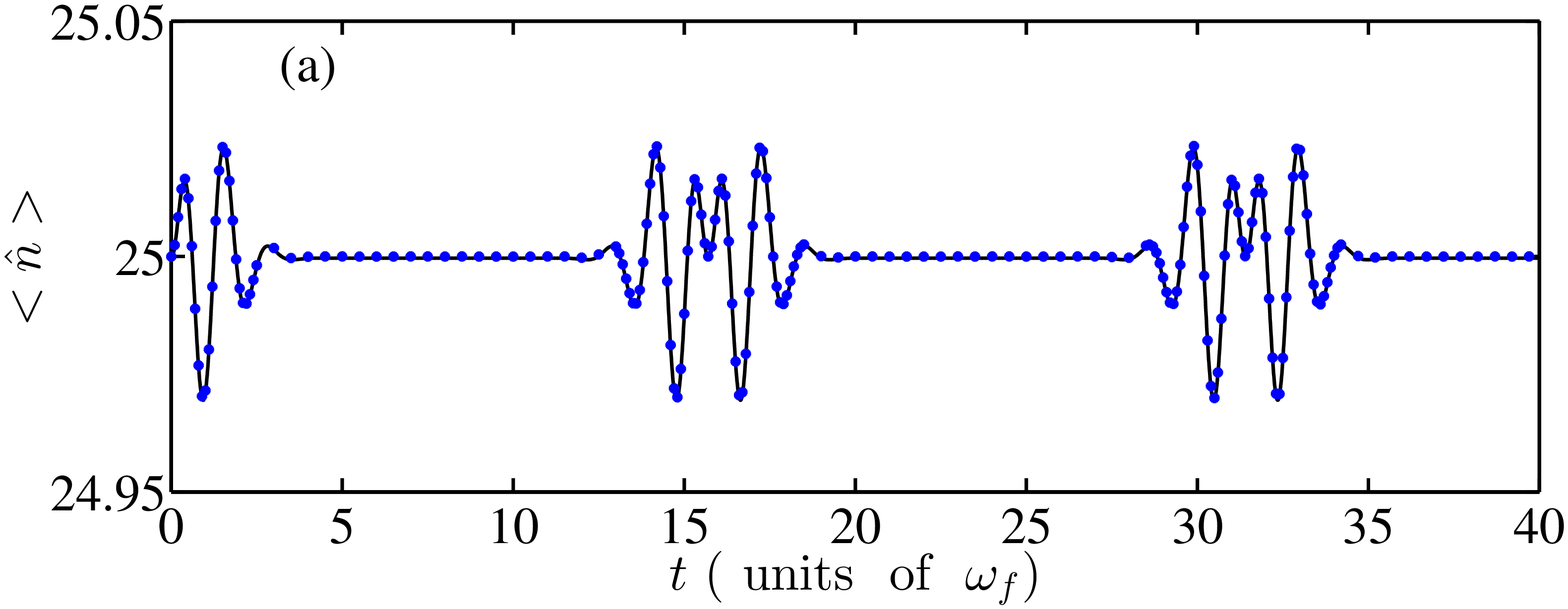}
\includegraphics[width=0.5\textwidth]{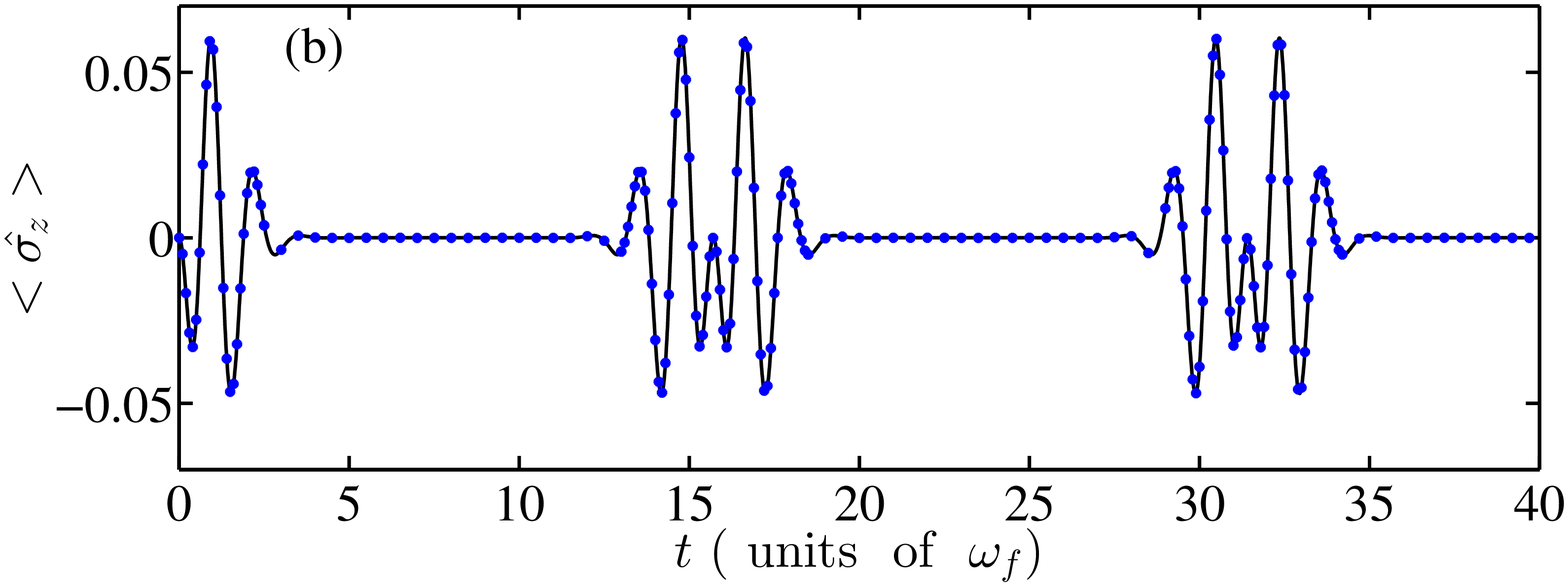}
\end{center}
 \caption{Evolutions of (a) mean photon number $\langle \hat{n} \rangle $ and (b) atomic inversion $ \langle \hat{\sigma}_{z} \rangle $ in the BS model with RWA. The initial state is the coherent state $ \vert t=0\rangle = \vert \alpha_{+},g\rangle + \vert \alpha_{-},e\rangle $ with $ \alpha=5 $ in the BS model when $ g_{+}=0 $ at resonance $ \omega_{f}=\omega_{0}=1$.  The coupling strength is $ g_{-}=0.1\omega_{f} $. For comparison, the results of Eqs.~(\ref{ite-10})--(\ref{Mt1}) (black line) and the TEEE method (blue points) are also plotted.
} \label{Fig1}
\end{figure}

We now apply the above numerical method to the BS model and compare the results with those of the TEEE method and Rodr\'{i}guez-Lara {\it et al.}~\cite{Rodrguez2013}. We first study the BS model with RWA ({\it i.e.}, $g_{+}=0$) at resonance, setting $ \omega_{f}=\omega_{0}=1 $. In this case, the BS model is exactly solvable. Figure~\ref{Fig1} shows the time evolutions of the mean photon number $\langle \hat{n} \rangle $ and the atomic inversion $ \langle \hat{\sigma}_{z} \rangle $, starting from a coherent initial state with $ \alpha=5 $.
Physically, the atomic inversion $ \langle \hat{\sigma}_{z} \rangle $ is equivalent to the atomic intensity difference between the two levels $\langle e \rangle $ and $\langle g \rangle $.
The atomic inversion collapses and revives with the variable photon number. Under the truncations $ N=20 $ and $ P=50 $, our results accorded with those of Rodr\'{i}guez-Lara {\it et al.}~\cite{Rodrguez2013} and the TEEE method (truncated with $P=60$). Similar agreement was observed in other parameter spaces (data not shown). Note that our method can be extended to very long-time evolution. Our tests confirmed highly accurate time-evolution behavior up to $t=600$ (in units of $\omega_f$), and the evolution time is extendible if necessary.

\section{BS model without RWA $(g_{+}\ne 0) $}
\label{sec:dis2}
\begin{figure}[htbp]
\begin{center}
 \centering
\includegraphics[width=0.5\textwidth]{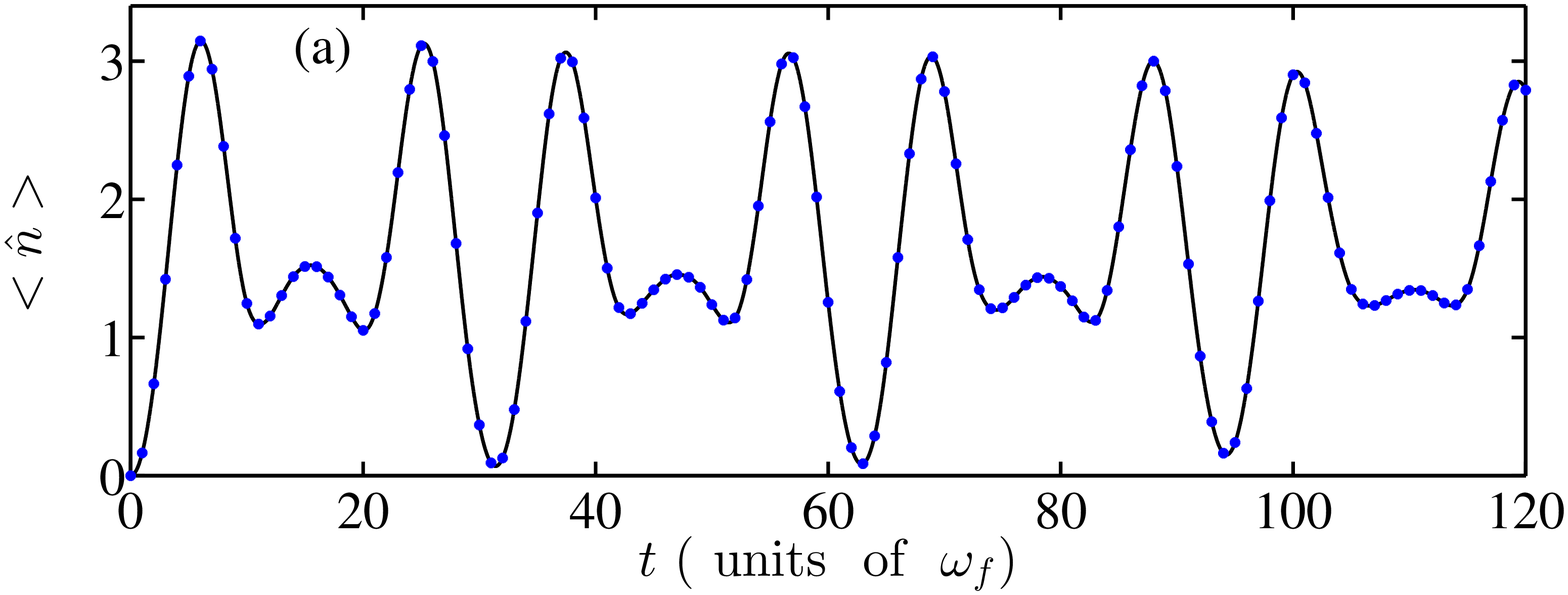}
\includegraphics[width=0.5\textwidth]{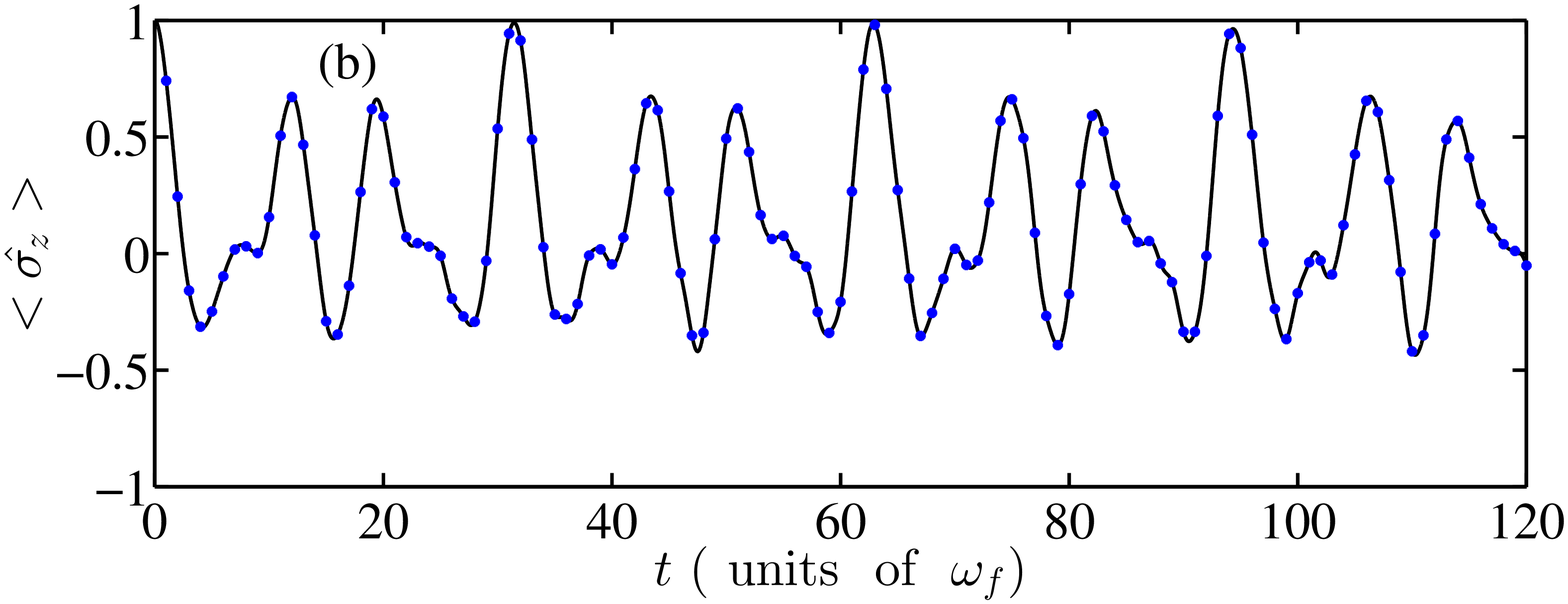}
\end{center}
 \caption{Evolution of (a) mean photon number $\langle \hat{n} \rangle $ and (b) atomic inversion $ \langle \hat{\sigma}_{z} \rangle $ in the RWA-void BS model. The initial state is a vacuum $ \vert t=0\rangle =\vert 0,e\rangle $ corresponding to $\alpha=0$ and $\theta=\pi/2$ in Eq.~(\ref{initialf}) (meaning that no photons are excited at $t=0$). Other parameters are $ \omega_{f}=1 $, $ \omega_{0}=3/4\omega_{f} $ and $ g_{-}=g_{+}=0.4\omega_{f} $. The results of Eqs.~(\ref{ite-10})--(\ref{Mt1}) (black line) and the TEEE method (blue points) are plotted for comparison.
 } \label{Fig2}
\end{figure}

In this section, we confirm that our method is also applicable to RWA-void BS model, {\it i.e.}, when $g_{+}\ne 0$. In this case, the BS model cannot be solved analytically. As before, the non-perturbation method is implemented by Eqs.~(\ref{ite-10})--(\ref{Mt1}). While our method must only recalculate $M(t)$ for the different $Q$s in Eq.~(\ref{Qg}), the TEEE method must numerically recalculate the eigenvalues and eigenvectors of the Hamiltonian.
Figure~\ref{Fig2} shows the evolutions of the atomic inversion $ \langle \hat{\sigma}_{z} \rangle $ and mean photon number $ \langle \hat{n} \rangle $ in RWA-void BS model, starting from the vacuum initial state $ \vert t=0\rangle = \vert 0,e\rangle $ with parameters $ \omega_{f}=1 $, $ \omega_{0}=3/4\omega_{f} $ and $ g_{-}=g_{+}=0.4\omega_{f} $ (corresponding to $\alpha=0$ and $\theta=\pi/2$ in Eq.~(\ref{initialf})).

Our results (truncated by $ N=30 $ and $ P=50 $) agree well with those of Rodr\'{i}guez-Lara {\it et al.}~\cite{Rodrguez2014} and the TEEE method (truncated by $P=60$),
demonstrating the applicability of our method to the RWA-void BS model. Note that the collapse and revival phenomena are absent in Fig.~\ref{Fig2}, as they were removed by the counter-rotating-wave term.

We subsequently consider the RWA-void BS model under the resonance condition $ \omega_{f}=\omega_{0}=1 $ with parameter values $ g_{-}=g_{+}=2\omega_{f} $. In this case, the Hamiltonian has no ground state and therefore has no physical meaning~\cite{Ng2000}. The aim is to confirm the behaviors of the mean atomic inversion $ \langle \hat{\sigma}_{z} \rangle $ and the mean photon number $\langle \hat{n} \rangle $, and the applicability of the numerical method. Rodr\'{i}guez-Lara {\it et al.}~\cite{Rodrguez2013} calculated the physics of this system via quantum optics methods. They assumed an analogy between the transport of single-photon states and the propagation of a classical field. The numerical propagation was stable on a 2000 $\times$ 2000 photonic lattice. However, the ground-state energy of the Hamiltonian is unbounded in this parameter space. In the na\"{\i}ve sense, this result implies an unstable mean atomic inversion and mean photon number. In the following, we first reproduce Rodr\'{i}guez-Lara {\it et al.'s} results by the TEEE method, then reveal the contradiction between the two arguments.

\begin{figure}[htbp]
\begin{center}
 \centering
\includegraphics[width=0.5\textwidth]{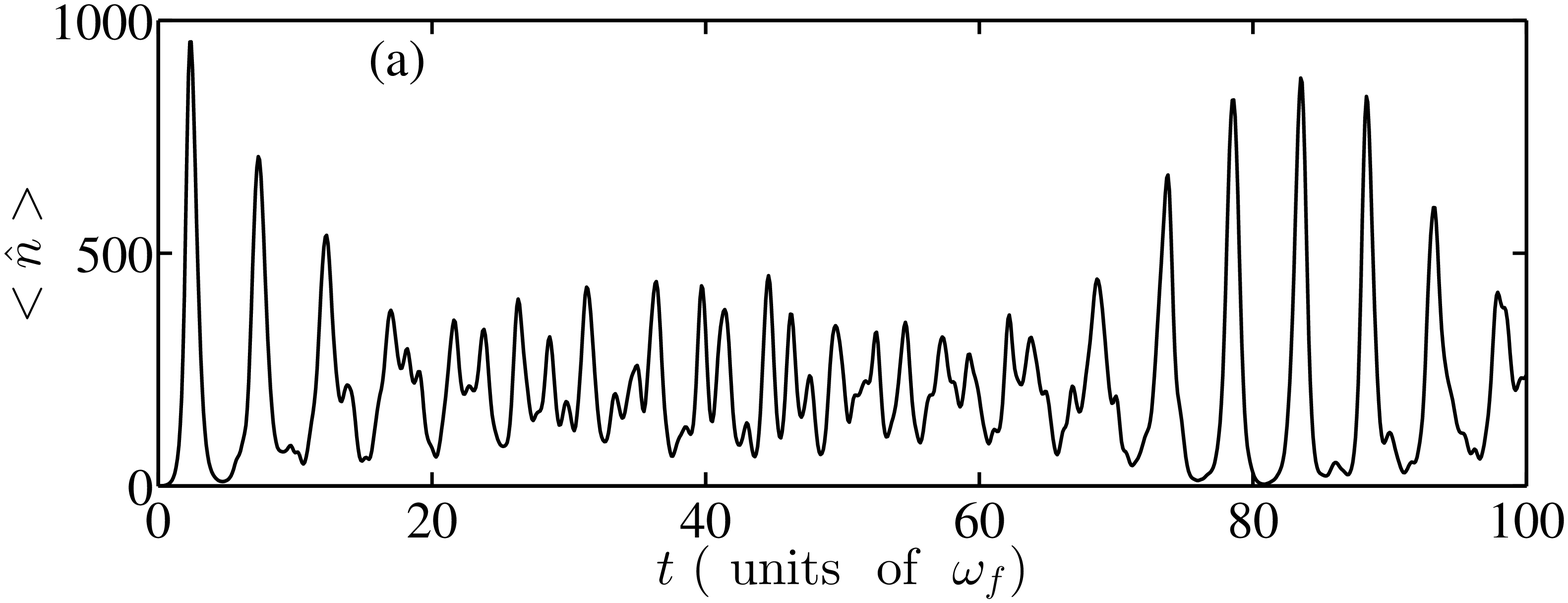}
\includegraphics[width=0.5\textwidth]{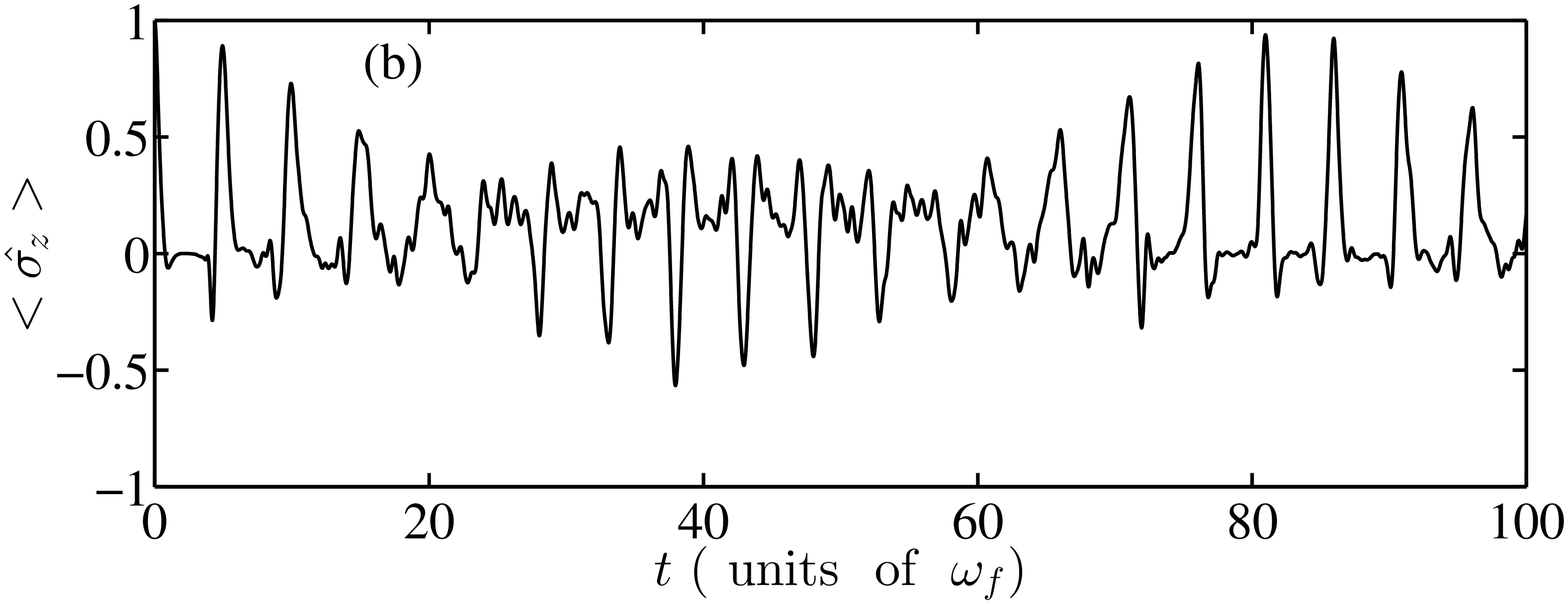}
\includegraphics[width=0.5\textwidth]{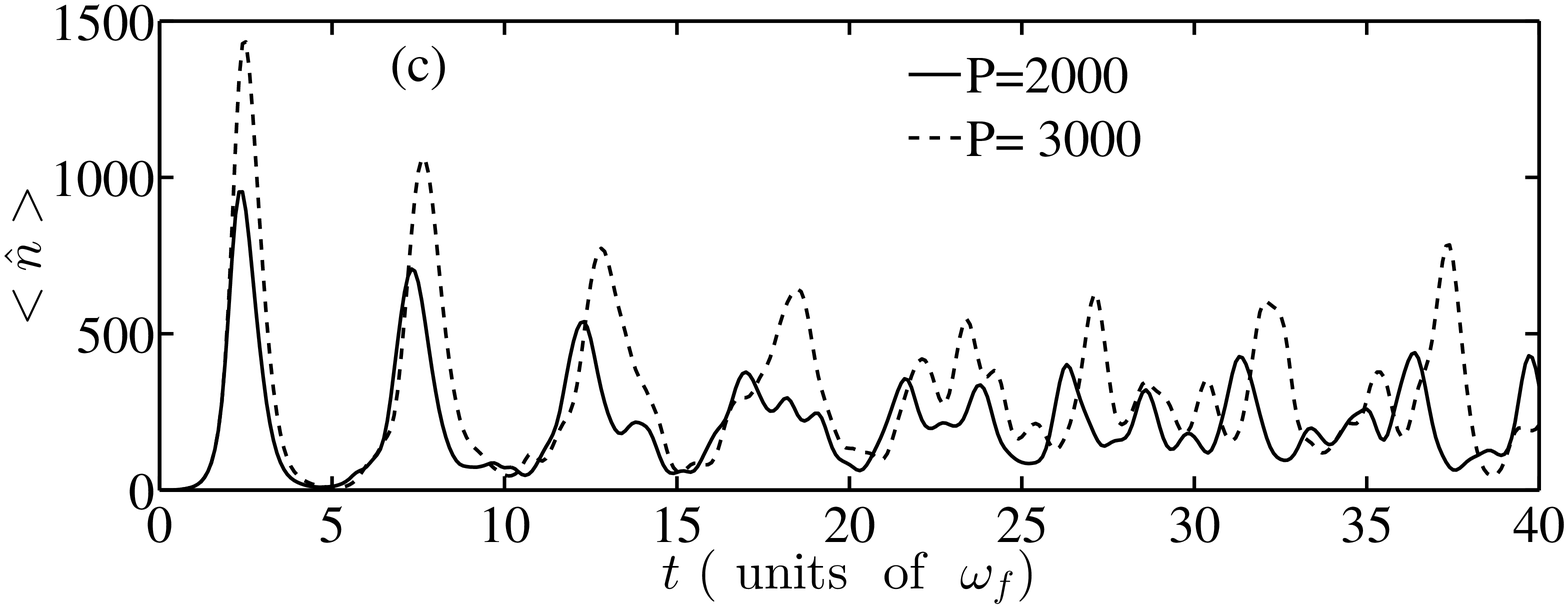}
\includegraphics[width=0.5\textwidth]{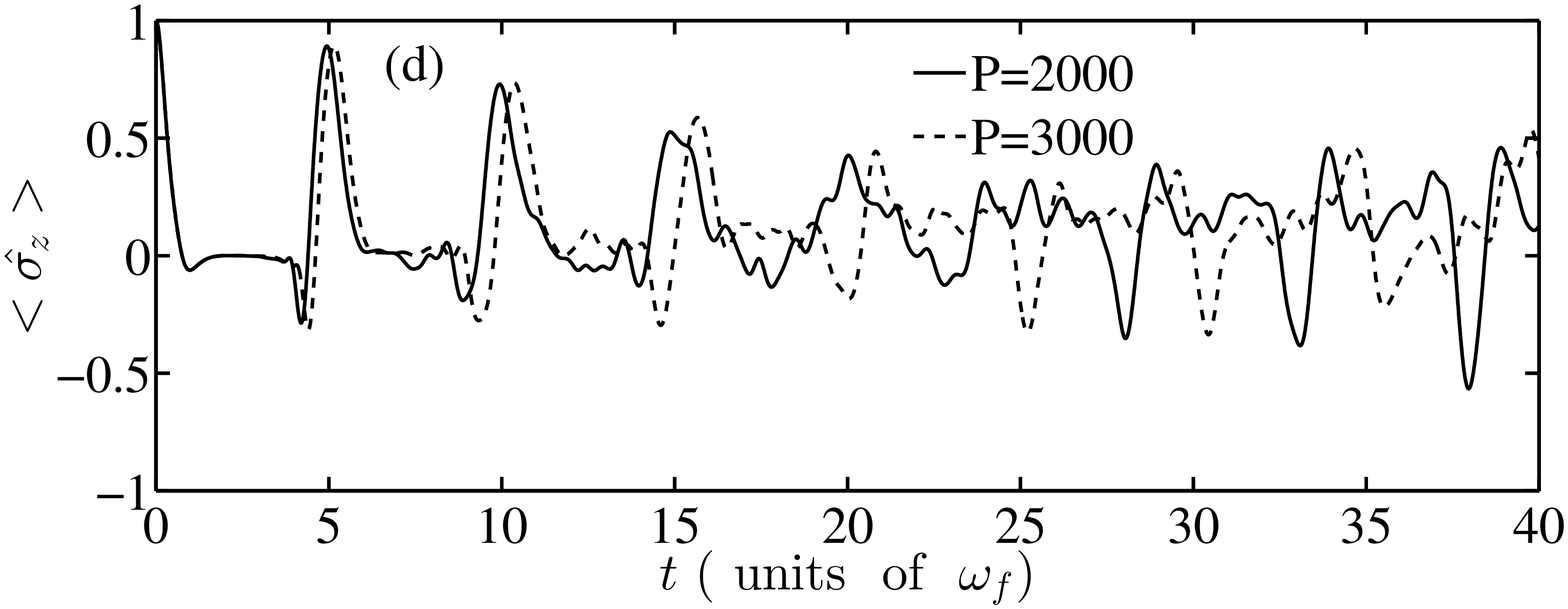}
\end{center}
 \caption{Evolutions of (a) mean photon number $\langle \hat{n} \rangle$ and (b) atomic inversion $ \langle \hat{\sigma}_{z} \rangle $ for $P=2000$. (c) and (d) represent the same data for (a) and (b), respectively, but for $P=2000$ and $3000$ simultaneously. The BS model begins from a vacuum state $ \vert t=0\rangle =\vert 0,e\rangle $ with $ \omega_{f}=\omega_{0}=1 $ and $ g_{-}=g_{+}=2\omega_{f} $.
 } \label{Fig3}
\end{figure}

In practice, Eq.~(\ref{Eq:TEEE}) is truncated by the limit of $P$ instead of being implemented in infinite dimensional Fock space. The numerical procedures of the TEEE method should therefore converge as the truncation number increases. In this case, the final results can be reliably obtained while satisfying the given precision requirements. As shown in panels (a) and (b) of Fig.~\ref{Fig3}, the mean photon number and mean atomic inversion results that were truncated by $P=2000$ matched those obtained by the TEEE method of Rodr\'{i}guez-Lara {\it et al.}~\cite{Rodrguez2013} with the same truncation factor.
However, when the truncation limit was increased to 3000 (see Fig.~\ref{Fig3}(c) and (d)), the results of our method and TEEE were very different (In order to see clearly the difference, we showed only the results for $t\in [0,40]$ comparing to Fig.~\ref{Fig3}(a) and (b)). Hence, whether the truncation requires to be increased or whether the discrepancy naturally arises from the non-physicality of the unbounded ground-state energy due to the incompleteness of the model must be elucidated in further analysis.
\begin{figure}[htbp]
\begin{center}
 \centering
\includegraphics[width=0.5\textwidth]{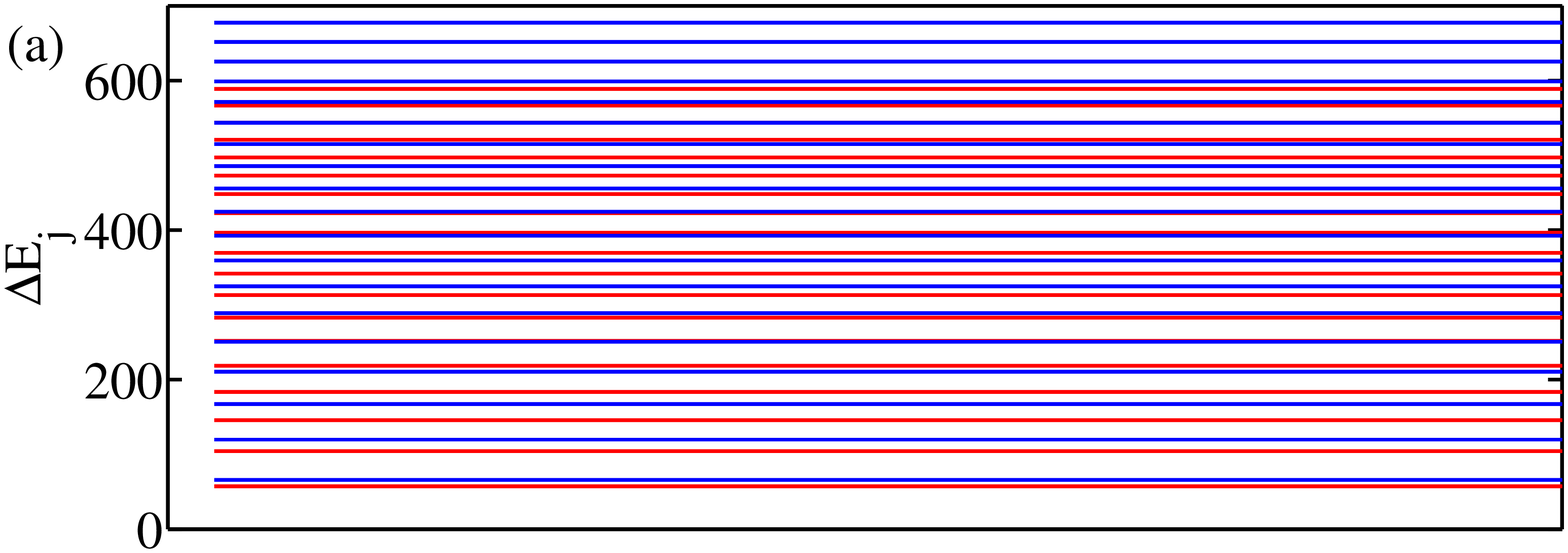}
\includegraphics[width=0.5\textwidth]{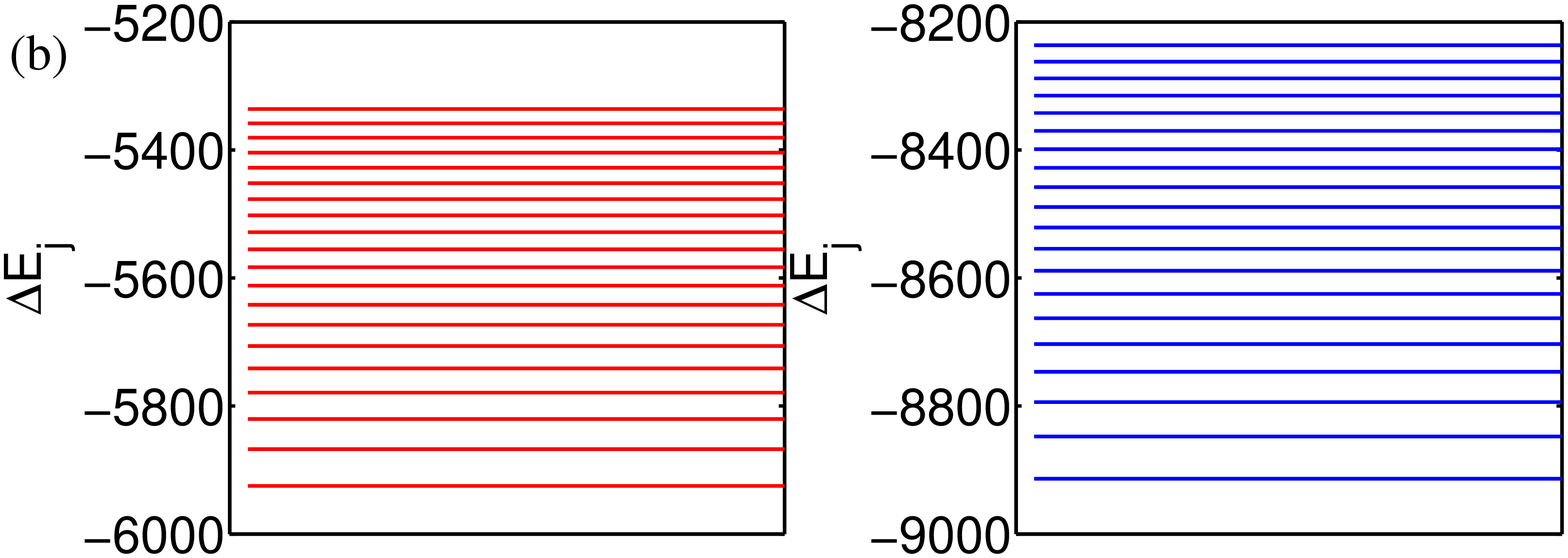}
\end{center}
 \caption{Top panel: Difference between the j-th ($ j=1,2,3,... $) excited-state energy and the ground-state energy ($\Delta E=E_{j}-E_{0} $) in the RWA-void BS model with parameters $ \omega_{f}=\omega_{0}=1 $ and $ g_{-}=g_{+}=2\omega_{f} $, calculated by the TEEE method truncated by $P= 2000$ (red line) and $P=3000$ (blue line). Bottom panel: Energy-level differences for $P=2000$ (red line in bottom left) and $P=3000$ (blue line in bottom right panel), respectively.
 } \label{Fig4}
\end{figure}

To resolve the above doubt, Fig.~\ref{Fig4}(a) shows the energy-level differences ($ \Delta E=E_{j}-E_{0} $) between the $j$-th ( $j=1,2,3,...$) excited state and the ground state in the RWA-void BS model. Figure~\ref{Fig4}(b) compares the ($ \Delta E=E_{j}-E_{0}$ in the models truncated by $P=2000$ and $P=3000$. Comparison of the two results clearly shows the invalidity of simply increasing the truncation limit, indicating that the non-convergence of the physical quantities originates from the incomplete Hamiltonian and the unbounded ground-state energy.

\begin{figure}[htbp]
\begin{center}
 \centering
\includegraphics[width=0.5\textwidth]{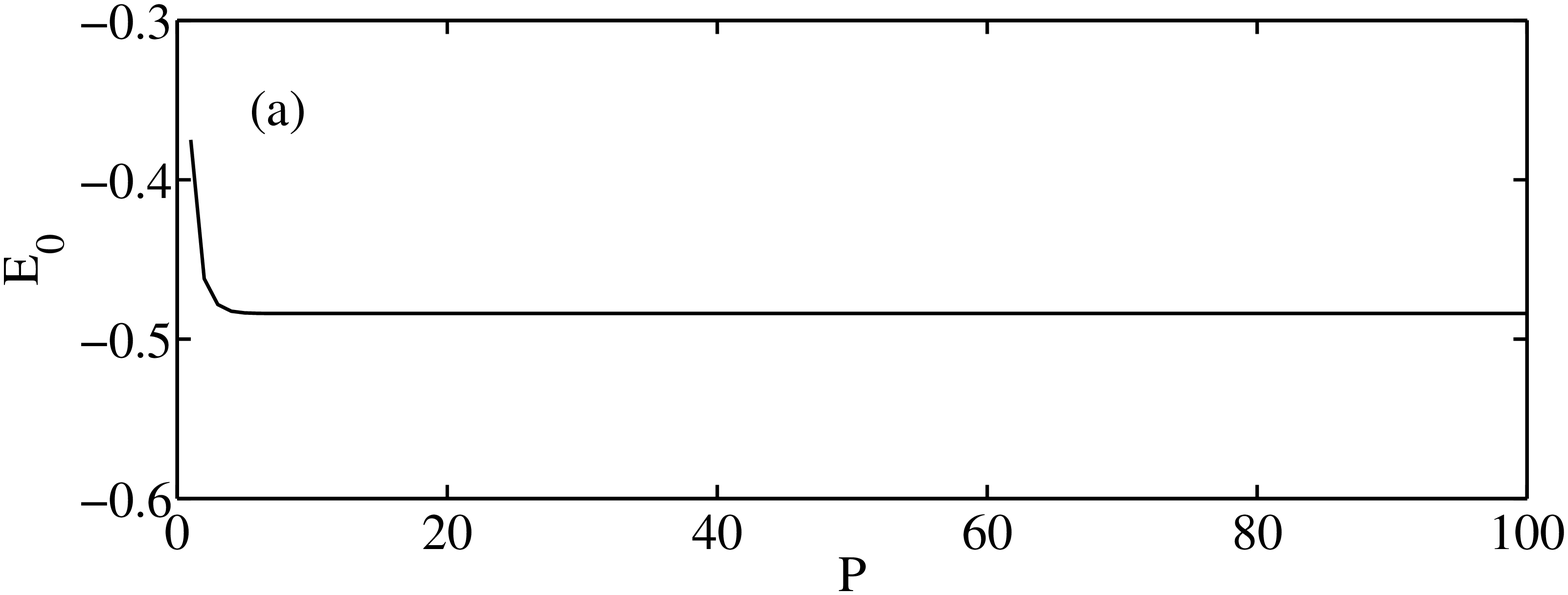}
\includegraphics[width=0.5\textwidth]{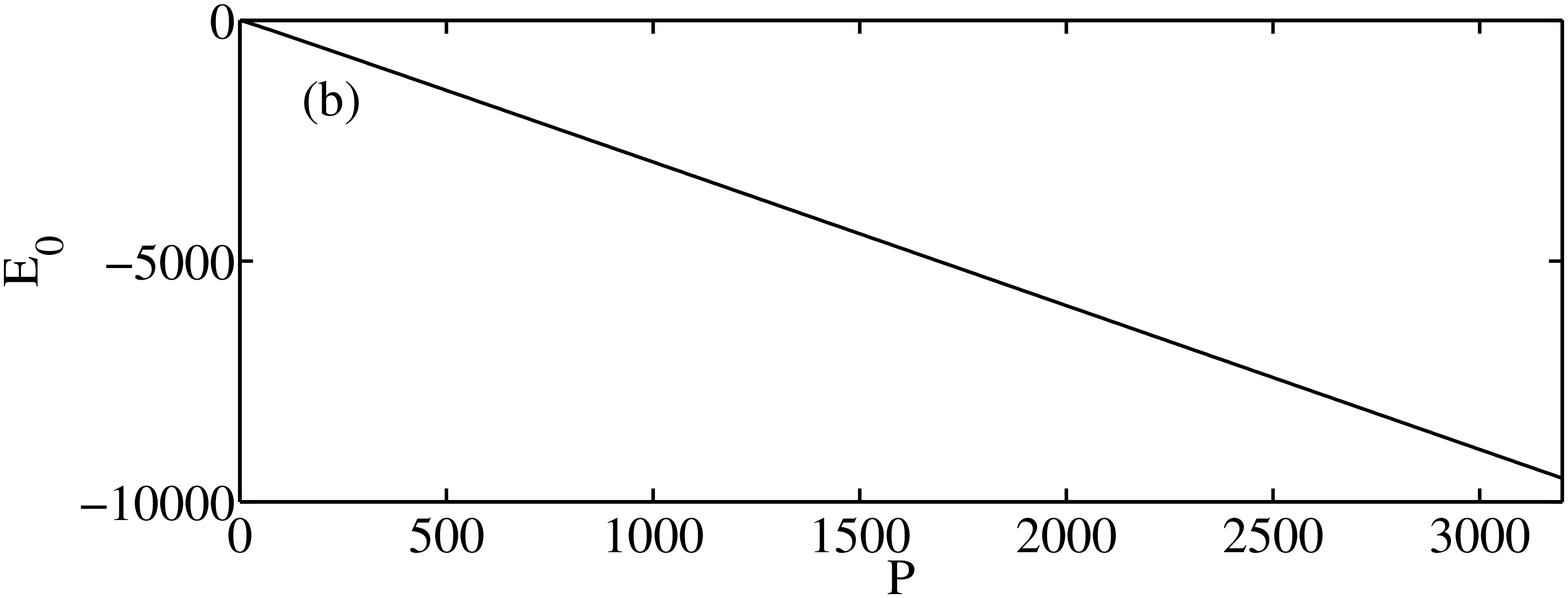}
\end{center}
 \caption{Ground-state energy versus truncation factor in the BS model with parameters (a) $ \omega_{f}=1 $ and $ \omega_{0}=3/4\omega_{f} $, $ g_{-}=g_{+}=0.4\omega_{f} $, and (b) $ \omega_{f}=\omega_{0}=1 $ and $ g_{-}=g{+}=2\omega_{f} $.
 } \label{Fig5}
\end{figure}

To further confirm this judgment, we studied the ground-state energy as a function of truncation. Figure~\ref{Fig5}(a) shows the results for $ g_{-}=g_{+}=0.4\omega_{f}$. After an initial rapid decrease, the ground-state energy stabilized at $P>6$.
Moreover, as shown in Fig.~\ref{Fig5}(b), the ground-state energy decreased linearly with increasing truncation limit for $ g_{-}=g{+}=2\omega_{f}$. In other words, the investigated system lacked a ground state and was physically infeasible.

\begin{figure}[htbp]
\begin{center}
 \centering
\includegraphics[width=0.5\textwidth]{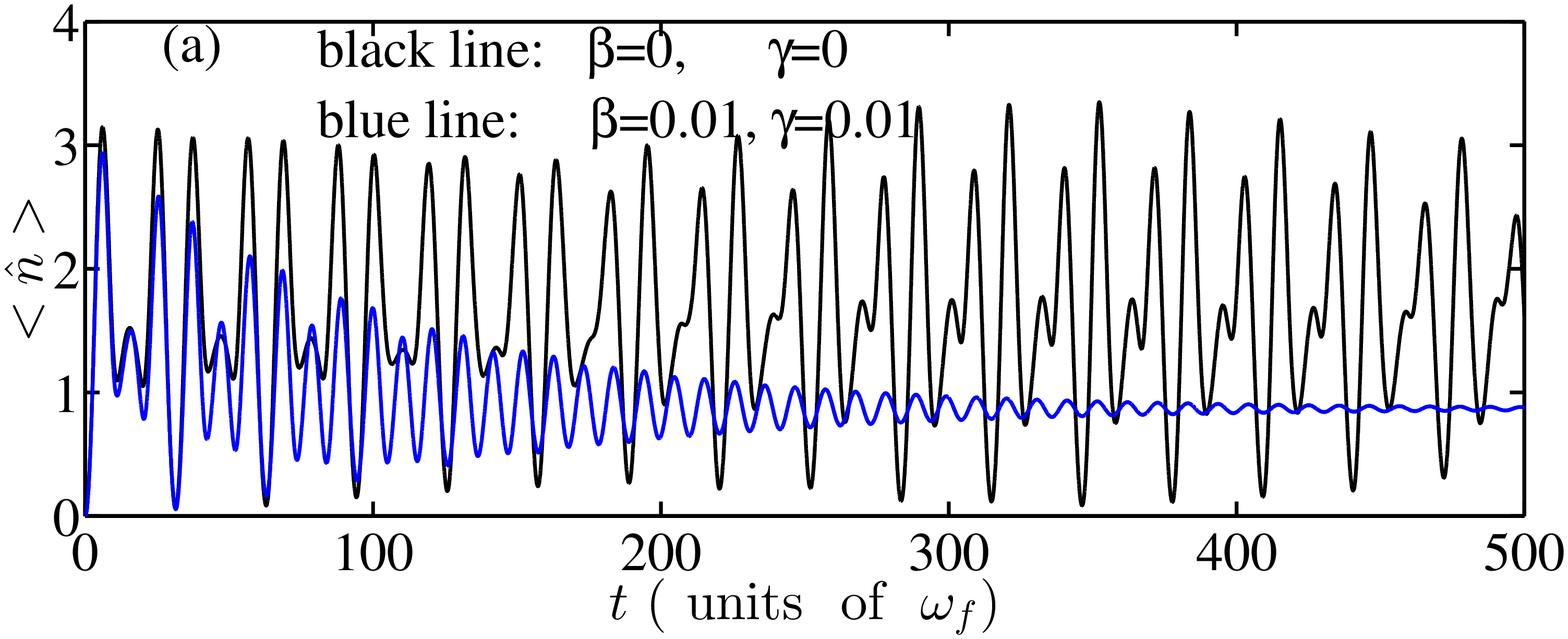}
\includegraphics[width=0.5\textwidth]{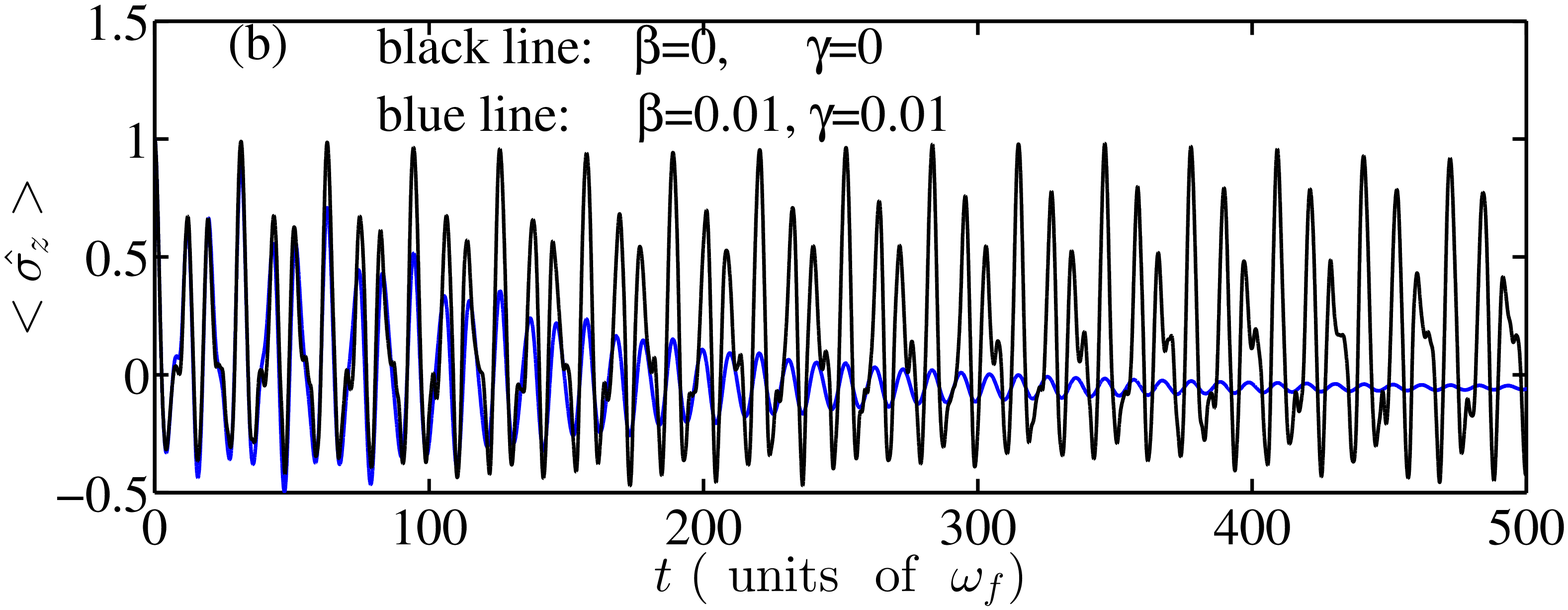}
\end{center}
 \caption{Evolutions of (a) mean photon number $\langle \hat{n} \rangle $ and (b) atomic inversion $ \langle \hat{\sigma}_{z} \rangle $ of the system for $ \beta=\gamma=0.01\omega_{f}$, starting from a vacuum initial state $ \vert t=0\rangle =\vert 0,e\rangle $. The other parameters are $ \omega_{f}=1 $, $ \omega_{0}=3/4\omega_{f} $ and $ g_{-}=g_{+}=0.4\omega_{f} $. For comparison, the results in the no-dissipation case $\beta=\gamma=0$ are plotted in black).
 } \label{Fig6}
\end{figure}

To incorporate the environmental influences in real-system dynamics, we must consider the irreversible processes. For this purpose, we included a purely imaginary coupling term comprising two phenomenological dissipative terms~\cite{non-Hermitian1} in the Hamiltonian~(\ref{H}) of the BS model as follows:
\begin{equation}
\hat{H}_{\rm dissipative}=\hat{H}-i\beta\hat{a^{+}}\hat{a}-i\gamma \vert e \rangle \langle e \vert,
\end{equation}
where $ \beta $ and $ \gamma $ are the photon leakage rate from the cavity and the spontaneous emission rate, respectively. The last (non-Hermitian) term renders the Hamiltonian as non-Hermitian. The reliability of the non-Hermitian Hamiltonian approach has been already tested in the dynamics of the pumped-dissipative JC model~\cite{non-Hermitian2}.
As our numerical method expands the initial state in Fock space, it is directly applicable to systems of dissipations. This property is absent in the TEEE method. Figure~\ref{Fig6} shows the evolution of the atomic inversion $ \langle \hat{\sigma}_{z} \rangle $ and the mean photon number $ \langle \hat{n} \rangle $ in this system, starting from a vacuum state $ \vert t=0\rangle = \vert 0,e\rangle $ (corresponding to $\alpha=0$ and $\theta=\pi/2$ in Eq.~(\ref{initialf})). Here, we set $ \beta=\gamma=0.01\omega_{f} $, and the other parameters as $ \omega_{f}=1 $, $ \omega_{0}=3/4\omega_{f} $, and $ g_{-}=g_{+}=0.4\omega_{f} $. Owing to finite damping of the photons and atoms, the oscillating photon number and atomic inversion in the dissipation case tended to steady values (see Fig.~\ref{Fig6}(a) and (b), respectively). The damping of the oscillations is attributable to interactions with the environment.

\section{\label{sec:level5}Conclusion}
\label{sec:con}
We proposed a systematic numerical method that calculates the time evolution of the physical quantities in the BS model of the interactions between a two-level atomic system and a field. Applying this method, we successfully transformed the Hamiltonian multiplication of $\hat{H}^n$ into a matrix multiplication of $Q^n$. This transformation greatly reduced the numerical calculation involved. Moreover, the matrix multiplication $Q^n$ is independent of the initial states, meaning that the time-dependent properties of the system can be computed for arbitrary initial states without recalculating the matrix multiplication $Q^n$.

The performance of the method was tested in numerical comparisons with the TEEE method. The method performed well in the whole coupling regime, and captured the collapses and revivals of the atomic inversion along with the photon-number variations. In the parameter space of a non-real positive-definite Hamiltonian, we also resolved the uncertainty regarding whether the evolution was stabilized by increasing the truncation limit over the dimension of the Fock space or whether it was a natural consequence of the unbounded ground-state energy (a nonphysical situation caused by the incompleteness of the Hamiltonian). We found that when the ground-state energy of the Hamiltonian is unbounded in a certain parameter space, the time evolution of the physical quantities is naturally unstable. Furthermore, the performance of the numerical method was tested in a dissipative BS model, which represents open quantum systems.

\begin{acknowledgments}
This work was supported by the NSF of China under Grant Nos. 11774316 and 11835011.
\end{acknowledgments}


\newpage
\small
\begin{thebibliography}{20}
\bibitem{JC1962} E.T. Jaynes and F.W. Cummings, Comparison of quantum and semiclassical radiation theories with application to the beam maser, Proc. IEEE {\bf \textbf{51}}, 89 (1963).

\bibitem{GB1991} J. Gea-Banacloche, Atom- and field-state evolution in the Jaynes-Cummings model for large initial fields, Phys. Rev. A {\bf \textbf{44}}, 5913 (1991).

\bibitem{RE2011} B. W. Shore?Sir Peter Knight and the Jaynes-Cummings model, J. Mod. Opt. {\bf \textbf{54}}, 2009 (2007).


\bibitem{HW2012} S. He, C. Wang, Q.H. Chen, X.Z. Ren, T. Liu, and K.L. Wang, First-order corrections to the rotating-wave approximation in the Jaynes-Cummings model, Phys. Rev. A {\bf \textbf{86}}, 033837 (2012).

\bibitem{Mb2015} M. Mirzaee and M. Batavani, Atom-field entanglement in the Jaynes Cummings model without rotating wave approximation, Chin. Phys. B {\bf \textbf{24}}, 040306 (2015).

\bibitem{FS2018} K. Fischer, S. Sun, D. Lukin, Y. Kelaita, R. Trivedi, and J. Vu\v{c}kovi\'{c}, Pulsed coherent drive in the Jaynes-Cummings model, Phys. Rev. A {\bf \textbf{98}}, 021802 (2018).

\bibitem{Rempe1987}
G. Rempe, H. Walther, and N. Klein, Observation of quantum collapse
and revival in a one-atom maser, Phys. Rev. Lett. {\bf \textbf{58}}, 353 (1987).

\bibitem{An1994}
 K. An, J.J. Childs, R.R. Dasari, and M.S. Feld, Microlaser: A laser with one atom in an optical resonator, Phys. Rev. Lett. {\bf \textbf{73}}, 3375 (1994).

\bibitem{Kundu2003}
A. Kundu, Quantum integrability and Bethe ansatz solution for interacting matter-radiation systems, J. Phys. A: Math. Gen {\bf \textbf{37}}, L281 (2004).

\bibitem{Braak2011}
D. Braak, Integrability of the Rabi model, Phys. Rev. Lett. {\bf \textbf{107}}, 100401 (2011).

\bibitem{MH2014}
L.J. Mao, S.N. Huai, and Y.B. Zhang, The two-qubit quantum Rabi model: Inhomogeneous coupling, J. Phys. A: Math. Theor. {\bf \textbf{48}}, 345302 (2015).

\bibitem{FS2014}
J.P. Feng, X.Z. Ren, and S. He, Entanglement dynamics of the Tavis-Cummings model without rotating wave approximation, Proc. SPIE {\bf \textbf{9269}}, 926916 (2014).

\bibitem{LL2018}
X.J. Liu, J.B. Lu, S.Q. Zhang, J.P. Liu, H. Li, Y. Liang, J. Ma, Y.J. Weng, Q.R. Zhang, H. Liu, X.R. Zhang, and X.Y. Wu, The Nonlinear Jaynes-Cummings Model for the Multiphoton Transition, Int. J. Theor. Phys. {\bf \textbf{57}}, 290 (2018).

\bibitem{A2012}
V.V. Albert, Quantum Rabi Model for N-state Atoms, Phys. Rev. Lett. {\bf \textbf{108}}, 180401 (2012).

\bibitem{Ng2000}
K.M. Ng, C.F. Lo, and K.L. Liu, Exact eigenstates of the intensity-dependent Jaynes-Cummings model with the counter-rotating term, Phys. A {\bf \textbf{275}}, 463 (2000).

\bibitem{BS1981}
B. Buck and C.V. Sukumar, Exactly soluble model of atom-phonon coupling showing periodic decay and revival, Phys. Lett. A {\bf \textbf{81}}, 132 (1981).

\bibitem{Celi2017}
A. Celi, A. Sanpera, V. Ahufinger, and M. Lewenstein,
Quantum optics and frontiers of physics: The third quantum revolution,
Phys. Scr. {\bf \textbf{92}}, 013003 (2017).

\bibitem{Wolf2013}
F.A. Wolf, F. Vallone, G. Romero, M. Kollar, E. Solano, and D. Braak, Dynamical correlation functions and the quantum Rabi model,
Phys. Rev. A {\bf 87}, 023835 (2013).

\bibitem{Xie2017}
Q.T. Xie, H.H. Zhong, M.T. Batchelor, and C. Lee, The quantum Rabi model: solution and dynamics, J. Phys. A {\bf 50}, 113001 (2017).

\bibitem{Cheng2017}
X.H. Cheng, I. Arrazola, J.S. Pedernales, L. Lamata, X. Chen, and E. Solano, Nonlinear quantum Rabi model in trapped ions, Phys. Rev. A {\bf \textbf{97}}, 023624 (2018).

\bibitem{Hu2017}
 B.L. Hu, H.L. Zhou, S.J. Chen, G. Xianlong, and K.L. Wang, Dynamical properties of the Rabi model, J. Phys. A: Math. Theor. \textbf{50}, 074004 (2017).

\bibitem{LW2009}
T. Liu, K.L. Wang, and M. Feng, The generalized analytical approximation to the solution of the single-mode spin-boson model without rotating-wave approximation, EPL {\bf \textbf{86}}, 54003 (2009).

\bibitem{ZX2014}
H. Zhong, Q. Xie, X. Guan, M.T. Batchelor, K. Gao, and C. Lee, Analytical energy spectrum for hybrid mechanical systems, J. Phys. A: Math. Theor. {\bf \textbf{47}}, 045301 (2014).

\bibitem{Rodrguez2013}
B.M. Rodr\'{i}guez-Lara, F. Soto-Eguibar, A.Z. C\'{a}rdenas, and H. M. Moya-Cessa, A classical simulation of nonlinear Jaynes-Cummings and Rabi models in photonic lattices, Opt. Express {\bf \textbf{21}}, 12888 (2013).

\bibitem{Cordeiro2008}
F. Cordeiro, C. Provid\^{e}ncia, J. da Provid\^{e}ncia, and S. Nishiyama, The Two-Photon Jaynes-Cummings Model: A Coherent State Approach, Adv. Studies Theor. Phys. {\bf \textbf{2}}, 181 (2008)
F. Cordeiro, C. Provid\^{e}ncia, J. da Provid\^{e}ncia, and S. Nishiyama, The Buck-Sukumar model described in terms of $ su(2)  \bigotimes  su(1, 1) $ coherent states, J. Phys. A: Math. Theor. {\bf \textbf{40}}, 12153 (2007).

\bibitem{Rodrguez2014}
B.M. Rodr\'{i}guez-Lara, Intensity-dependent quantum Rabi model: spectrum, supersymmetric partner, and optical simulation, J. Opt. Soc. Am. B. {\bf \textbf{31}}, 1719 (2014).

\bibitem{non-Hermitian1}
A. Sergi and K.G. Zloshchastiev, Non-Hermitian quantum dynamics of a two-level system and models of dissipative environments, Int. J. Mod. Phys. B {\bf \textbf{27}}, 1350163 (2013);
A. Sergi and K.G. Zloshchastiev, Time correlation functions for non-Hermitian quantum systems, Phys. Rev. A {\bf \textbf{91}}, 062108 (2015).

\bibitem{non-Hermitian2}
S. Echeverri-Arteagaa, H. Vinck-Posadaa, E.A. G\'{o}mez, A comparative study on different non-Hermitian approaches for modeling open quantum systems, arXiv: 1810.11376;
S. Echeverri-Arteaga, H. Vinck-Posada, and E.A. G\'{o}mez, A comparative study on the reliability of non-Hermitian effective Hamiltonian approach for modeling open quantum systems, Optik {\bf \textbf{171}}, 413 (2018).
\end{thebibliography}
\end{document}